\documentclass[preprint,12pt]{elsarticle}

\usepackage{epsfig}
\usepackage{array,tabularx,epsfig,mathrsfs,graphicx,rotating}
\usepackage{ifthen}
\usepackage{amsfonts}
\usepackage{ragged2e}
\usepackage{verbatim}

\newcommand{\beq}{\begin{equation}}
\newcommand{\eeq}{\end{equation}}

\chardef\til=126

\journal{Comp. Phys. Communication}

\begin{document}
\begin{frontmatter}

\title{ProMC:  Input-output data format for HEP applications using varint encoding}

\vspace{2.5cm}

\author[lab1]{S.V.~Chekanov\corref{cor1}}
\ead{chekanov@anl.gov}
\author[lab1]{E.~May}
\author[lab2]{K.~Strand}
\author[lab1]{P.~Van Gemmeren}

\address[lab1]{HEP Division, Argonne National Laboratory,
9700 S.Cass Avenue,
Argonne, IL 60439
USA
}

\address[lab2]{
Department of Physics, Winona State University, 175 West Mark St.,~Winona, MN 55987, USA
}

\cortext[cor1]{Corresponding author}

\begin{abstract}
A new data format for Monte Carlo (MC) events, or any structural data,
including experimental data, is discussed.  The format is designed to store data
in a compact binary form using variable-size integer encoding
as implemented in the Google's Protocol Buffers package.
This approach is implemented in the {\sc ProMC} library which
produces smaller file sizes for MC records compared to the
existing input-output libraries used in high-energy physics (HEP).
Other important features of the proposed format are
a separation of abstract data layouts from concrete programming implementations, self-description
and random access. Data stored in {\sc ProMC} files
can be written, read and manipulated
in a number of programming languages, such  C++, JAVA, FORTRAN and PYTHON.
\end{abstract}


\end{frontmatter}


\section{Introduction}

A crucial requirement for many scientific applications is to store, 
retrieve and process large-scale numeric data with a small signal and large 
background (or ``noise''). Information on background objects is not required to be 
stored with the same relative numeric precision as that for signal objects. 
An input/output library which dynamically streams data depending on the content of information 
becomes important for effective data storage and analysis. 

	A typical example is  the Large Hadron Collider (LHC) experiments designed to 
investigate  proton-proton and heavy-ion collisions in order to understand the basic structure of matter. 
The LHC experiments are currently involved in event processing and physics analysis of petabytes of data.
A single analysis requires a processing of tens of terabytes of data 
located on the grid storage across the globe. 
The  number of collisions recorded by the ATLAS experiment since 2009 exceeds 20 billion. 
The number of particles in a single collision  will increase by a factor 5-10 for future high-luminosity LHC runs. 
Currently, the LHC experiments store more than 100 petabytes of data and this number 
will increase by a factor 10 over the next 10 years. 
Most stored data has  
a small fraction of ``signal'' particles, while most of low-energetic particles from other events
are less interesting and represent ``pileup'' background.
It is important to store pileup particles to derive corrections, 
but to store such particles with the same numeric precision as signal particles is not justified and inefficient.

This paper discusses an input-output library which has a content-dependent compression of data for files. 
It stores less energetic particles with reduced relative numeric precision and smaller numbers of bytes compared to
to high-energetic particles. 
The library is designed for Monte Carlo (MC) simulation events,
but it can naturally be extended to store any information. The library was created during the Snowmass Community Studies \cite{snowmass}
with the goal to store MC simulations in a compact form on public web pages.

\section{The proposal}
This paper discusses an input/output persistent  framework which:

\begin{itemize}
\item
streams data into a binary form and dynamically writes less interesting, 
low-energetic particles with a reduced numeric precision compared to more energetic ``signal'' particles. 
For example, it is expected 
that such content-dependent compression  may decrease the LHC data volume by 30\% or more \cite{promc}.
Although we use the word ``compression'', it should be noted that no
standard compression algorithms (gzip, zip, bunzip2) are used since 
the file-size reduction is achieved using a highly efficient binary format.

\item
is multiplatform. Data records can be manipulated in C++, JAVA and PYTHON. This opens the possibility 
to use a number of ``opportunistic'' platforms for data analysis, such as Windows or Android,  which have not been used widely 
in HEP.

\item
is a self-describing data format based on a template approach to encode complex data structures. 
One can generate C++, JAVA and PYTHON analysis codes from the  file itself. 

\item
has random access capabilities. Events can be read starting at any index. Individual events can be accessed 
via the network without downloading the entire files. 
Metadata information can be encoded for each record, allowing for a fast access to interesting events.

\item
is implemented as a simple, self-containing library which can easily be deployed on a 
number of architectures including
supercomputers, such as IBM Blue Gene/Q system. 
\end{itemize}

The proposed input-output framework is expected to be used in many scientific areas. In particular, it is useful for
(a) data reduction for large general-purpose detectors at colliders and other experiments; 
(b) effective data preservation due to small file sizes, backward compatibility and self-descriptive property; 
(c) effective data analysis without CPU overhead due to the standard decompression algorithms.

\section{Existing approaches}

The LHC experiments store data and experiment-specific MC events in compressed ROOT format \cite{root}. 
To store events generated by MC models in a more generic and exchangeable way, HEPMC \cite{hepmc}, STDHEP \cite{stdhep},
HepML \cite{Belov:2010xm} and the Les Houches event format (LHEF) \cite{Alwall:2006yp} 
file formats have been developed. 
For example, the  {\sc HEPMC} library is interfaced with all major Monte Carlo models and is widely used by the HEP community
due to its simplicity,
platform independence, exchangeability and reusability.
However, the HEPMC format stores data in  uncompressed ASCII files, 
which are typically ten times larger than  ROOT files with the default compression. 

        The ROOT IO is an integrated part of the C++ ROOT analysis framework \cite{root} developed at CERN.
This framework is heavily integrated in the Linux platform.
It uses the  ``gzip'' compression which is a CPU intensive and lacks 
flexibility for storing particles depending on their importance.
As discussed before,  the current and future LHC experiments will collect events with only a small fraction of
signal particles that are important for analyzers,  while most of low-energetic particles
from other (overlayed) events represent ``pileup'' background.
For high-luminosity LHC runs, one ``signal'' event (for example, event with a Higgs particle) will
contain 50-140 pileup events, with up to 10,000 low-energy particles that are not 
important for analysis of signal signatures. Still,  such particles (or a fraction of such particles) 
should be kept to derive corrections to signals.
Therefore, to store low-energetic particles with a smaller numeric precision becomes crucial in effective data
storage and analysis. The fixed-number of bytes to represent numeric data used by ROOT and by other data formats
does not allow implementation of a compression that depends on particle properties
(particle energy, mass, origin, etc.).

\section{Varint data encoding}

A possible solution for data reduction is to use
``varints'' \cite{protobuf} which can encode integer ({\tt int32}, {\tt int64}, etc.) values using variable number of bytes\footnote{
In the case of 4-momentum, one can convert a float variable to varint using a predefined conversion 
factor}.
Such algorithm is implemented in the Google's Protocol Buffers library \cite{protobuf}.
This library encodes complex data in the form of platform-neutral ``messages''.
A message is a  logical record of information containing a series of name-value pairs. 
Smaller integer numbers represented by varints in such messages use a smaller number of bytes compared to large numbers. 
For HEP applications, this implies that four-momenta of low-energetic particles  encoded using the integer values
can be represented with a smaller number of bytes.
In addition, many particle
characteristics (such as particle status, particle ID, etc.) should be  represented by integer values 
anyway and this is  well suited to the varint representation.

Historically, the  approach to store HEP data using Google's Protocol Buffers  was first attempted in the 
 {\sc JHepWork}\footnote{{\sc JHepWork} was renamed to {\sc SCaVis} \cite{scavis} in 2013.} 
data-analysis framework \cite{Chekanov:1261772} in 2008, which  
offered the {\sc CBook} C++  package to keep Monte Carlo records and other structural data using varints. 
Later, the {\sc Protocol Buffers} library became the core  of another HEP library, the so-called A4 project \cite{2012JPhCS.396b2012E},
that had the goal of providing fast I/O  for structured data. 

	Although the varint data encoding is available in the {\sc Protocol Buffers} library 
publicly released by Google, this library alone is not sufficient to pursuit the goal of 
creating large files with multiple logically-separated records. 
The Protocol Buffers approach is most effective if each separate Protocol Buffers message has a
size of less than 1~MB (as recommended by Google). 
The  major problems that need to be addressed are: 1) to design Protocol Buffers message to store
particles in a single event to allow for the varint representation; 2) to find a method of serialization
of multiple messages (``events'') into a file which can keep many events; 3) how to implement metadata
model for fast access of interesting events and particles.
While (1) is rather specific to HEP, (2) and (3) are very general issues
that have to be solved in any research area where logically-separated event
records with varint-based  information is an attractive option for data
storage and processing. Because of such problems,
the usage of Google's Protocol Buffers to keep large numeric data is still limited in science and technology.

\section{Current implementation}
\label{cimp}

The following  sections will discuss the current implementation of the library, called {\sc ProMC}, which
implements all the features discussed above. In the following, we will use  small-caps typeface fonts to indicate 
the {\sc ProMC} library implemented in C++,
while files generated using this library, ``ProMC'' files, will be shown using the normal fonts. 
A similar convention is applied for other library names, such as 
the {\sc Protocol Buffers} library that is used to write and read Protocol Buffers messages.

\begin{figure}
\begin{center}
\includegraphics[width=0.28\textwidth]{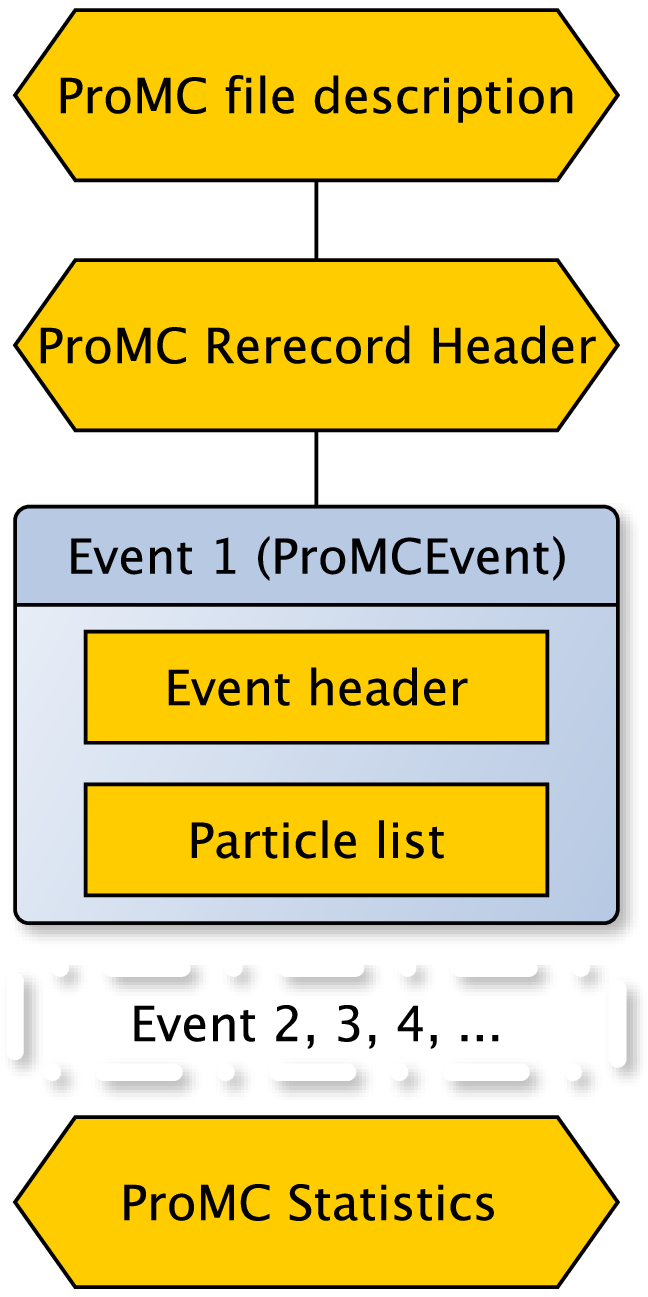}
\end{center}
\caption{
A schematic representation of the ProMC file format. 
All records are encoded using the Protocol Buffers messages.
In addition, some metadata information is stored as text files inside the ZIP file for easy access on platforms without 
the installed {\sc ProMC} library.
}
\label{fig:graph}
\end{figure}

The {\sc ProMC} C++ library \cite{promc} is designed to store HEP collision events
using the Google's {\sc Protocol Buffers} library as a backend. 
The data are  stored in a file with the file headers and
multiple logically-separated messages.
Each separate message leverages the varint encoding for representing a single MC event.    
Figure~\ref{fig:graph} shows a schematic representation of a ProMC file.
All Protocol Buffers  messages are stored as ZIP entries inside the ProMC file using
the {\sc Zipios++} package \cite{Zipios} for reading and writing ZIP files through the standard C++ iostreams.
The ZIP method for archiving supports lossless data compression, as implemented in the {\sc ZLIB} library \cite{zlib}.
However, this library is
only used to organize binary Protocol Buffers's messages (which do not require compression) in the ProMC files.
In this sense, ZIP is a method of archiving binary records, rather than the actual method
of compressing event records. 
The ZIP compression, however, is used for some metafile records, such as text templates
describing file layouts and logfiles which can be embedded inside the ProMC files. 

The {\sc Protocol Buffers} library (version 2.5) is included in the {\sc ProMC} package to avoid 
clashes  with the already installed {\sc Protocol Buffers} library (which can be version 2.42  for many Linux distributions),
to provide better self-containment and to simplify the deployment of examples and conversion tools
which use a predefined location of the {\sc Protocol Buffers} library.
However, {\sc ProMC} can also be installed using the existing {\sc Protocol Buffers} library, as described
in the ProMC web page \cite{promc}. In this form, all ProMC Makefiles of the conversion tools
should be redesigned.

To work with the ProMC files, the {\sc ProMC} C++ library does not need to be installed. 
This C++ library has to be installed
if events will be written or read in C++. ProMC files can also be read and created using JAVA or PYTHON,
without the platform-dependent {\sc ProMC} library.

The current {\sc ProMC} library is built on the assumption that the new  data format should be self-describing
and can generate analysis source codes (in C++, JAVA, PYTHON) from a ProMC file itself without knowing how
it was originally created. 
Additional notable features of {\sc ProMC} are random access to any given event and a possibility to stream 
individual events through the network without reading or downloading entire files. 

Several benchmarks have shown that ProMC files are rather compact, typically 
40\% smaller than ROOT files assuming {\tt Double32\_t} types  for float values and the default ROOT compression.
Table~\ref{xtab1} shows the file sizes for 10,000 $t\bar{t}$ events generated with  {\sc PYTHIA 8} \cite{Sjostrand:2007gs} 
for a $pp$ collider at $14$~TeV. The ProMC files are $38\%$ smaller than files with the same information using  
ROOT, and significantly smaller 
than LHEF \cite{Alwall:2006yp} and HEPMC (production release: 2.03.11) \cite{hepmc} files, 
including those with the compression based on the gzip, bzip2 and lzma algorithms.
In case of events with large pileup (i.e. a large fraction of soft particles), ProMC files
can lead to  almost a factor two smaller files compared to ROOT files with the same information
stored using {\tt Double32\_t} data types.

The same table shows benchmark tests for the read speed. For these tests, 
the files were opened, and all entries with 4-momenta of particles were extracted, but no  calculations are performed. 
The read speed of the ProMC files is $30\%$ faster than for the ROOT files based on the default gzip
compression,  
and is substantially faster compared to the other formats. 
It can be seen that this difference in the read speed is roughly proportional to file sizes. 
Files in the LHEF and HEPMC  formats after compression were not tested,  
given that such, technically challenging tests, can hardly show much improvement 
in the read speed since a typical time for file decompression is 1-2 min.
No significant difference in the file creation speed was detected (this test was dominated by event generation). 

In addition to the C++ compiled programs, benchmark tests were also performed using programs implemented in JAVA and JYTHON (the PYTHON language implemented in JAVA) 
using the {\sc SCaVis} \cite{scavis,Chekanov:1261772} framework. The JAVA Virtual Machine (JVM) processes the ProMC files faster
than the programs implemented in C++.  
This indicates that  JVM creates a more optimized binary code to deliver a better performance.  
In the case of LHEF ASCII files and JVM, the benchmark program parses all lines and tokenizes the strings, without attempting to build complete particle record, therefore, such
test may not be accurate when comparing with the ROOT and ProMC approaches.
 
Benchmarks tests using PYTHON implemented in C have also been performed.
The read speed of ROOT files (67 sec) was found to be substantially faster than for the ProMC files (980 sec).
This was explained by the fact that the PYTHON benchmark program for the ProMC file tests  was implemented in pure PYTHON 
and does not use a C++ binding (unlike PyROOT that uses C++ libraries). In order to perform a fair
comparison for the PYTHON benchmark programs, a C++ backend for the {\sc Protocol Buffers} should be enabled.
This has not been implemented yet for the {\sc ProMC} library.

The ProMC file size  depends on many factors, but there are three major factors that should be mentioned:
 1)  energy distributions of stored particles; 2) what information is stored; 3) how the information 
is represented using integer values and  a typical range of  
integer values. 
In the first case, events with large fraction of low-energy (``soft'') particles will use smaller file size than those 
that contain particles with large values of 4-momenta of particles. 
The file size depends on a conversion factor which converts float values to integer representation.

The default ProMC mapping between the energy units used in HEP and integer values is given in Table~\ref{xtab2} for
a typical $pp$ collision experiment  at centre-of-mass (CM) energies up to $20$ TeV. 
This mapping between integer type in C++/JAVA and the varint type was obtained by 
multiplying energies, masses and 4-momentum components
expressed in  GeV by the factor $10^{5}$, and rounding the resulting value to nearest integer.    
For a collider at $\sqrt{s}=100$~TeV, this mapping can fail for storing 
particles (jets) with energies above $21$~TeV,
therefore, a multiplication factor should be reduced to $10^{4}$ in order to be store possible particles or jets close to 
this CM energy range. On the other hand, the conversion factor can be increased for low-energy experiments. 
To avoid overflows for large energies, 
the primary rule to remember when using the multiplicative conversion factor is that $\sqrt{s}$ times
the multiplicative factor should not be larger than $2^{31}-1$ assuming integer data type is in the 32-bit representation.
Another consideration is the required relative numeric precision, since large
numbers may lead to less effective storage for varints.

Table~\ref{xtab2a} shows the default mapping 
between energy  (or masses,4-momentum etc.) values and the 32-bit integer representation using the varint encoding.
The last column shows the rounding errors which should be noted when restoring the data.
This table corresponds to the default conversion factor $10^{5}$ as discussed above and which is
suitable for the current LHC experiments.
Table~\ref{xtab2b} shows a suggested conversion and relative rounding errors using the conversion
factor $10^{4}$  for post-LHC era colliders.

The varint conversion factor for float values is included in the metadata section of the ProMC file format.
It should be restored while processing ProMC files.
This approach provides  a certain flexibility:  
For example, instead of using a single constant conversion factor that leads to relative numerical
precision that changes with the stored energy, one can use an energy-dependent conversion factor that leads
to roughly constant, energy-independent, relative rounding error. 
It should also be pointed out that the {\sc ProMC} library can store particle's 4-momenta 
using float or double precision types, while keeping varints only for information that requires 
integer values. 
In this case, the ProMC data-size  reduction will be less  
effective.

Another factor that determines data reduction depends on the fraction of information which can be represented 
using integer types, such as  {\tt int32} and {\tt int64}.
For a typical MC truth event record, several particle characteristics can be written using 
varints without loosing numeric precision. Such examples include particle ID,
status code, 1st and 2th daughter and mother particles. In many cases, their values are small 
and thus can be represented by only a few bytes using {\tt int32}.

The current approach to the data reduction by particle experiments is often based 
on removing particles (or tracks, calorimeter cells, etc.) below some  transverse-momentum cut, which is typically  0.2-0.5~GeV for the LHC experiments. 
Such rejection of the information on final-state particle collisions may not be required for the ProMC files that use a smaller number of bytes
to store low-energetic particles.

\subsection{Working with ProMC files}
\label{tools}

The {\sc ProMC} package can be downloaded from the HepForge web page \cite{promc}.
The only required external library is {\sc ZLIB} \cite{zlib} which is used to 
organize individual binary messages with separate MC events as discussed in Sect.~\ref{cimp}.

After installation as discussed in the online manual, an environmental variable ``PROMC'' pointing 
to the installation directory should be defined. 
The installed package contains the directory ``\$PROMC/examples'' with a number 
of examples, converters from other formats, as well as  with several tools   
to work with the ProMC  files.

Table~\ref{xtab3} lists the tools included in the {\sc ProMC} package to work with ProMC files. 
These programs
are implemented in PYTHON and stored in the directory ``\$PROMC/bin''.  

As discussed before, the ProMC files are simple ZIP archives, thus they can be unzipped to view
the file structure. This can be observed without unpacking individual records as  ``{\tt  unzip -l <ProMC file>}''. 
A proper unpacking the ProMC file can be done with the same command but without the option "-l".
Each file represents an event record written as the binary file that can be read using the {\sc Protocol Buffers} library. 

The ZIP program can also be used to extract any given event. For example, extracting an event 100 and saving it to a 
file "100.event" will require to run: 

\begin{verbatim}
unzip -p <ProMC file>  100 > 100.event
unzip -p <ProMC file>  ProMC.proto  > ProMC.proto
\end{verbatim} 
The last example creates a text file that describes the Protocol Buffers platform-neutral layout of the event. 
Similarly, one can look at the attached logfile and print the number of stored events as:

\begin{verbatim}
unzip -p <ProMC file>  logfile.txt 
unzip -p <ProMC file>  promc_nevent 
\end{verbatim}
In these examples, we send the contents of the files "logfile.txt" and "promc\_nevent" via pipe into a Linux 
shell console.

Table~\ref{xtab4} lists  the converters from/to ProMC format included with the {\sc ProMC} package. 
These programs are located in the directory ``\$PROMC/examples'' and should be compiled 
for final deployment. 

The ProMC files can also be written from FORTRAN programs using an external package called {\tt FortranProMC}.
This package should  be downloaded
and compiled separately, as long as the original {\sc ProMC} package is installed.
The {\tt FortranProMC} includes an example that shows how to fill
ProMC files using the {\it PYTHIA6} \cite{Sjostrand:2006za} generator.
This example can be used to create ProMC files using any FORTRAN-based generator.

\subsection{Reading ProMC files}

Since ProMC files are self-describing, one  can generate analysis codes in C++, JAVA, PYTHON
from the available file, even without knowing how data are organized inside the file.
For this example, the C++ {\sc ProMC}  library should be installed.

\begin{verbatim}
promc_info   <name>.promc # check information
promc_proto  <name>.promc # extracts Protocol Buffers files 
promc_code                # create analysis code in C++, JAVA, PYTHON
make                      # compiles the C++ code
\end{verbatim}
This example shows how to access the information about the existing file using the command
{\tt promc\_info}.  Next, the command {\tt promc\_proto} extracts the platform neutral description of data
layout in the Protocol Buffers format. Such description files will be located in the directory ``proto'' and can be used
to generate analysis code for reading or writing ProMC files.  Finally,
the command {\tt promc\_code}  generates the analysis codes in C++, PYTHON and JAVA. Such files will be located in the corresponding
directories where this command is executed. The execution of the ``make'' compiles the C++ analysis code.

\subsection{ProMC browser} 

ProMC files can be accessed without any external C++ library using a  
browser implemented in JAVA. The browser, together with the complete source code, 
is included in the directory ``examples/browser''  of the {\sc ProMC} package which can be downloaded 
separately from the ProMC web page \cite{promc}. 

\begin{figure}
\begin{center}
  \includegraphics[width=1\textwidth]{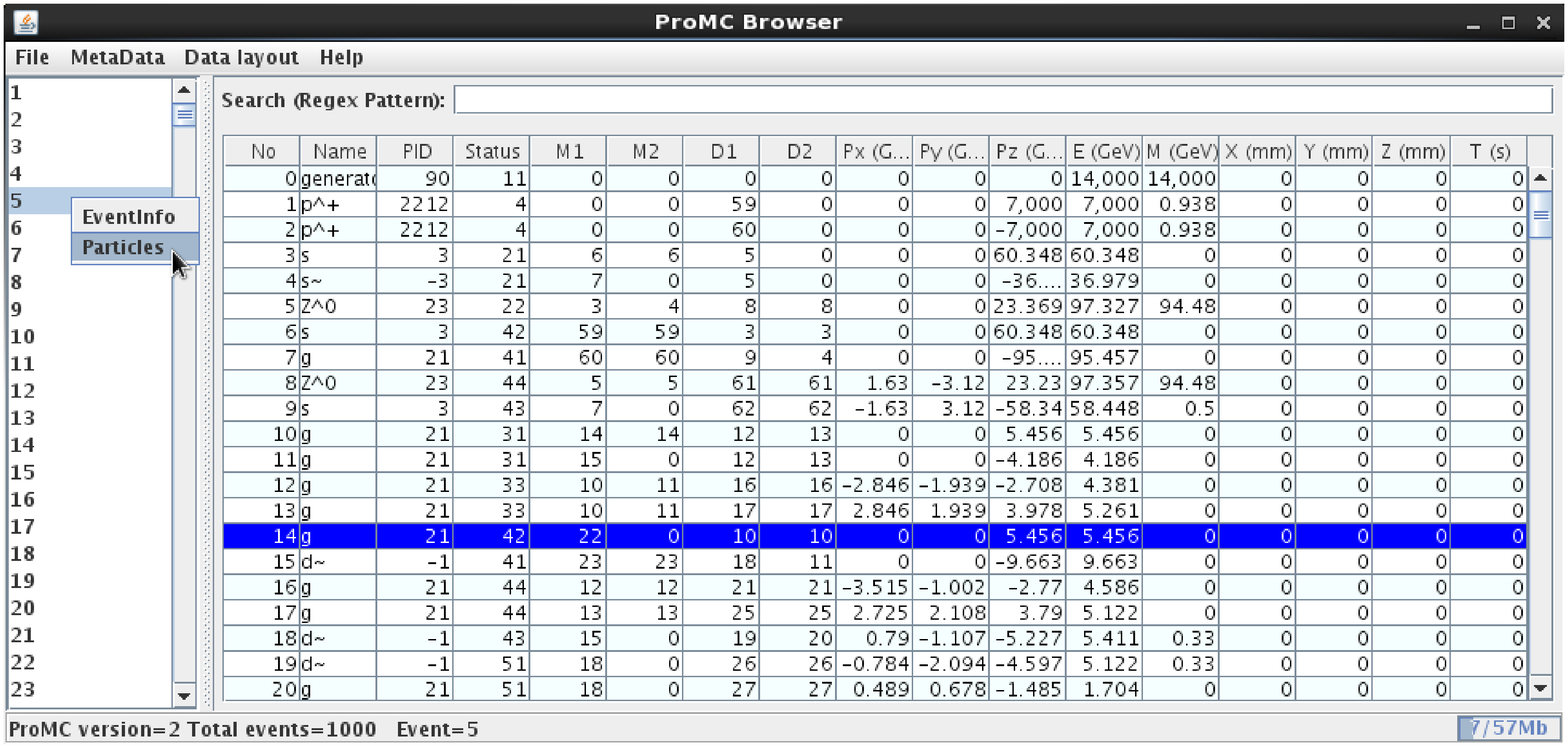}
  \caption{A JAVA browser for ProMC files with events from a MC generator. The upper field is a search for a given particle name.}
  \label{fig:browser}
\end{center}
\end{figure}

Figure \ref{fig:browser} shows the browser window with particle records for a specific event. 
The browser also displays  general information about the ProMC file as well as the
metadata file included in the ProMC header. 
The browser can be accessed by either of 
the following commands:

\begin{verbatim}
java -jar browser_promc.jar <name>.promc  (or) 
promc_browser <name>.promc 
\end{verbatim}
where {\tt <name>.promc} is a ProMC file with the extension {\tt .promc}. 

A ProMC file can also be accessed via URL links  without the need to download the file on the hard drive. 
This can be done by revising the above commands to  the following:

\begin{verbatim}
java -jar browser_promc.jar <file URL> (or) 
promc_browser <file_URL>
\end{verbatim}

The above example can be used to view event records from leading-order parton-shower MC simulations.
Next-to-leading order event generators typically have a few particles from hard interactions,
but the event records contain additional information on event weights. In this case, one should use
the following syntax:

\begin{verbatim}
java -cp browser_promc.jar probrowser.NLO <file URL> (or)
promc_browser_nlo <name>.promc 
\end{verbatim}
Unlike leading-order parton-shower Monte Carlo models, this brings up a window which can be used to view an array of weights
representing uncertainties on predictions.
More information about how to store events in the ProMC files 
from NLO event generators is given in Ref.~\cite{Chekanov:2014fgas}. 

The ProMC browser can also display information on separate events, platform-independent 
data layout files and logfiles  (if such files are embedded into the ProMC file structure).

\section{ProMC usage and documentation}

The {\sc ProMC} library was used for the Snowmass 2013 community studies in order to 
store truth and {\sc DELPHES} \cite{Ovyn:2009tx,deFavereau:2013fsa} MC events in a compact binary form on the 
web servers for HEP community
The {\sc DELPHES} version 3.10 fast simulation \cite{deFavereau:2013fsa} can read the ProMC files 
with the truth records and convert
such files to reconstructed events. 
Due to its compactness and fast read speed, {\sc ProMC} is used for the     
{\sc HepSim} repository \cite{Chekanov:2014fgas} with data from theoretical computations. This file repository  
includes MC events from leading-order parton-shower generators as well as weighted events from  next-to-leading order  QCD calculations.

Since {\sc ProMC} is implemented as a simple, self-containing library, it was deployed on a
number of platforms including high-performance computers, such as IBM BlueGene/Q, located at the
Argonne Leadership Computing Facility, the description of which is beyond the scope of this paper.
The {\sc ProMC} library is installed on CERN AFS and, at this moment, under tests within the   ATLAS collaboration.

A number of examples illustrating how to read and write the ProMC files is given on the {\sc ProMC} web page \cite{promc}.
The web page includes several examples of how to read, write and manipulate with ProMC files 
using several programming languages: C++, JAVA and PYTHON. Some basic information on stored data can also be extracted using PHP and other
languages that can use the {\sc Protocol Buffers} library.

For C++ and PYTHON examples of reading and writing data, ROOT/PyROOT can be used for graphical visualization.
There are also examples which illustrate 
how to read data using JYTHON, the PYTHON language implemented in JAVA. In this case, no platform specific libraries are used to read and display the data.
In  case of JAVA or JYTHON, {\sc SCaVis} \cite{scavis,Chekanov:1261772} and {\sc Jas} \cite{java_toni} JAVA-based analysis environments 
can be used for visual representation of data (histograms, scatter plots etc.). 
There is also an example which shows the random access capability of the ProMC format 
and how to access certain events from a network without reading the entire ProMC file.
In addition, a few examples are given which illustrate how to fill ProMC
 files directly from the {\sc PYTHIA 8} \cite{Sjostrand:2007gs} MC generator or from HEPMC
files.

\subsection{Current limitations}
\label{limitations}

The current prototype version of the {\sc ProMC} library is 1.3. 
The file-size limitation of this version for both the ProMC 1.3  files and Protocol Buffers messages
is given by the {\sc ZLIB} archive library, which is $2^{32}-1$ bytes (or 4 GB minus 1 byte).
In order to handle larger files, other libraries supporting  ZIP archives should be implemented.
It should be pointed out that the 4~GB restriction does not limit the usage of the current {\sc ProMC} prototype library for
production:
The information stored in a  4~GB ProMC file is equivalent to that stored in a $32$~GB of uncompressed LHEF, and such large LHEF files are unpractical
for real usage. 

Generally, there is no limit on the  number of files stored in the ProMC files. The ProMC files can hold
any number of events that can be read by PYTHON or JAVA. 
However, it was observed that the {\sc Zipios++} \cite{Zipios} package used for the standard C++ iostreams  
has a limit of 65520 entries. 
This limitation is a consequence of the chosen {\sc Zipios++} library to read ProMC entries in C++ programs, rather than 
a principle limitation of the {\sc ProMC} format. This problem will be corrected in future.

\section{Conclusion}

The {\sc ProMC} C++ library \cite{promc}  is available for download and testing.
The current version of this library is 1.31. Although it is an early-stage prototype, the {\sc ProMC} library has already 
been used in a number of projects
as discussed in this paper. 
ProMC has a potential to be an important  file format for current and future experiments
since it leads to small file sizes which are suitable  for effective data storage, has fast data access,
and supports multiple programming languages.

\section*{Acknowledgements}
One of us (S.C.) would like to thank J.~Proudfoot for a discussion.
The submitted manuscript has been created by UChicago Argonne, LLC,
Operator of Argonne National Laboratory (``Argonne'').
Argonne, a U.S. Department of Energy Office of Science laboratory,
is operated under Contract No. DE-AC02-06CH11357.
This research used resources of the Argonne Leadership Computing Facility at Argonne National Laboratory, which is supported by the Office of Science of the U.S. Department of Energy under contract DE-AC02-06CH11357.

\newpage
\bibliographystyle{elsarticle-num}
\bibliography{biblio}

\begin{thebibliography}{10}
\expandafter\ifx\csname url\endcsname\relax
  \def\url#1{\texttt{#1}}\fi
\expandafter\ifx\csname urlprefix\endcsname\relax\def\urlprefix{URL }\fi
\expandafter\ifx\csname href\endcsname\relax
  \def\href#1#2{#2} \def\path#1{#1}\fi

\bibitem{snowmass}
{HEP Community Summer Study (Snowmass)}, \url{http://www.snowmass2013.org/}
  (2013).

\bibitem{promc}
S.~Chekanov, {Next generation input-output data format for HEP using Google's
  protocol buffers}, contributed Papers Submitted to the Snowmass 2013 Study.
  SNOW13-00090. Project URL: \url{http://promc.hepforge.org/} (2013).
\newblock \href {http://arxiv.org/abs/1306.6675} {\path{arXiv:1306.6675}}.

\bibitem{root}
I.~Antcheva, et~al., {ROOT: A C++ framework for petabyte data storage,
  statistical analysis and visualization}, Comput. Phys. Commun. 180 (2009)
  2499--2512.
\newblock \href {http://dx.doi.org/10.1016/j.cpc.2009.08.005}
  {\path{doi:10.1016/j.cpc.2009.08.005}}.

\bibitem{hepmc}
M.~Dobbs, J.~Hansen, L.~Garren, L.~Sonnenschein, {{HEPMC User Manual}},
  \url{http://lcgapp.cern.ch/project/simu/HepMC/}.

\bibitem{stdhep}
L.~Garren, P.~Lebrun, {{StdHep. A common output format for Monte Carlo
  events}}, \url{http://cepa.fnal.gov/psm/stdhep/}.

\bibitem{Belov:2010xm}
S.~Belov, L.~Dudko, D.~Kekelidze, A.~Sherstnev, {HepML, an XML-based format for
  describing simulated data in high energy physic}, Comput.~Phys.~Commun. 181
  (2010) 1758.
\newblock \href {http://arxiv.org/abs/1001.2576} {\path{arXiv:1001.2576}},
  \href {http://dx.doi.org/10.1016/j.cpc.2010.06.026}
  {\path{doi:10.1016/j.cpc.2010.06.026}}.

\bibitem{Alwall:2006yp}
J.~Alwall, A.~Ballestrero, P.~Bartalini, S.~Belov, E.~Boos, et~al., {A Standard
  format for Les Houches event files}, Comput.~Phys.~Commun. 176 (2007) 300.
\newblock \href {http://arxiv.org/abs/hep-ph/0609017}
  {\path{arXiv:hep-ph/0609017}}, \href
  {http://dx.doi.org/10.1016/j.cpc.2006.11.010}
  {\path{doi:10.1016/j.cpc.2006.11.010}}.

\bibitem{protobuf}
{Google}, Protocol buffers. google's data interchange format,
  \url{http://code.google.com/apis/protocolbuffers/} (2008).

\bibitem{scavis}
S.~Chekanov, {{SCaVis. Scientific Computation and Visualization Environment}},
  \url{http://jwork.org/scavis/} (2013).

\bibitem{Chekanov:1261772}
S.~Chekanov, Scientific data analysis using Jython Scripting and Java,
  Springer-Verlag, London, 2010, iSBN 978-1-84996-286-5, e-ISBN
  978-1-84996-287-2.

\bibitem{2012JPhCS.396b2012E}
J.~{Ebke}, P.~{Waller}, {The A4 project: physics data processing using the
  Google protocol buffer library}, Journal of Physics Conference Series 396~(2)
  (2012) 022012.
\newblock \href {http://arxiv.org/abs/1208.1600} {\path{arXiv:1208.1600}},
  \href {http://dx.doi.org/10.1088/1742-6596/396/2/022012}
  {\path{doi:10.1088/1742-6596/396/2/022012}}.

\bibitem{Zipios}
{{Zipios++. A library for reading and writing Zip files using standard C++
  iostreams}}, \url{http://zipios.sourceforge.net/} (2013).

\bibitem{zlib}
G.~Roelofs, M.~Adler, {ZLIB 1.2.8. A compressing file-I/O Library)},
  \url{http://www.zlib.net/} (1996).

\bibitem{Sjostrand:2007gs}
T.~Sjostrand, S.~Mrenna, P.~Z. Skands, {A Brief Introduction to PYTHIA 8.1},
  Comput.~Phys.~Commun. 178 (2008) 852.
\newblock \href {http://arxiv.org/abs/0710.3820} {\path{arXiv:0710.3820}},
  \href {http://dx.doi.org/10.1016/j.cpc.2008.01.036}
  {\path{doi:10.1016/j.cpc.2008.01.036}}.

\bibitem{Sjostrand:2006za}
T.~Sjostrand, S.~Mrenna, P.~Z. Skands, {PYTHIA 6.4 Physics and Manual}, JHEP 05
  (2006) 026.
\newblock \href {http://arxiv.org/abs/hep-ph/0603175}
  {\path{arXiv:hep-ph/0603175}}.

\bibitem{Chekanov:2014fgas}
S.~Chekanov, {HepSim: a repository with predictions for high-energy physics
  experiments}, project URL: \url{http://atlaswww.hep.anl.gov/asc/hepmc/}
  (2014).
\newblock \href {http://arxiv.org/abs/1403.1886} {\path{arXiv:1403.1886}}.

\bibitem{Ovyn:2009tx}
S.~Ovyn, X.~Rouby, V.~Lemaitre, {DELPHES, a framework for fast simulation of a
  generic collider experiment}, Tech. rep. (2009).
\newblock \href {http://arxiv.org/abs/0903.2225} {\path{arXiv:0903.2225}}.

\bibitem{deFavereau:2013fsa}
J.~de~Favereau, C.~Delaere, P.~Demin, A.~Giammanco, V.~Lemaitre, et~al.,
  {DELPHES 3, A modular framework for fast simulation of a generic collider
  experiment} (2013).
\newblock \href {http://arxiv.org/abs/1307.6346} {\path{arXiv:1307.6346}}.

\bibitem{java_toni}
A.~Johnson, A java based analysis environment {JAS},
  \url{http://jas.freehep.org/jas3/} (1996).

\end{thebibliography}

\newpage
\section*{Tables}

\begin{table}[h] 
    \begin{tabular}{lcccc}
    \hline
    File format  & File size (in MB)  & \multicolumn{3}{|c}{Read speed (in seconds)} \\ \hline
                 &                    &  C++  &  JAVA VM  &  JYTHON \\ \hline\hline
     {\sc ProMC}  &  307              &  15.8 &   11.7,12.1* & 33.3, 34.6*   \\
     {\sc ROOT}  &  423               &  20.4 &   --         &  --           \\
     {\sc LHEF}  &  2472              &  84.7 &   9.0, 9.6*  &  --          \\
     {\sc HEPMC} &  2740              &  175.1 &  --         &  --       \\
     {\sc LHEF} (gzip)  &  712        &  --    &   --        &  --          \\
     {\sc HEPMC}(gzip)  &  1021       &  --    &   --        &  --          \\
     {\sc LHEF} (bzip2)  &  552        &  --    &   --        &  --          \\
     {\sc HEPMC}(bzip2)  &  837       &  --    &   --        &  --          \\
     {\sc LHEF} (lzma)  &  513        &  --    &   --        &  --          \\
     {\sc HEPMC}(lzma)  &  802        &  --    &   --        &  --          \\
     \hline
    \end{tabular}
\caption{Typical file sizes for 10,000 $t\bar{t}$ events generated for a $pp$ colliders at $\sqrt{s}=14$~TeV. The table
also shows the read speed using a C++, JAVA and JYTHON (the Python language implemented in JAVA and running inside the JVM).
The ROOT uses {\tt Double32\_t} for float values and the default compression.  
For all tests, the memory cache on Linux was cleared to avoid the data caching.
The programs were tested  on Intel(R) Xeon(R) CPU X5660 @ 2.80GHz.
In case of C++, the benchmark program reads complete particle records using the appropriate ROOT or {\sc ProMC} file libraries.
The numbers indicated with asterisks give the read speed for tests that take into accoung the 
intialization of the JVM before file processing.
In the case of LHEF file format and JAVA benchmark, the program parses all lines and tokenizes the strings, 
without attempting to build  MC particle records, therefore, this test may not be accurate.
The read speed for the test programs implemented in PYTHON (written in C) is discussed in the text. 
}
\label{xtab1}
\end{table}

\begin{table}
    \begin{tabular}{llc}
    \hline
    Energy  &  integer representation & Nr of bytes  \\ \hline
    0.01 MeV      & 1  & 1   \\
    0.1  MeV      & 10  & 1   \\
    1    MeV      & 100  & 2   \\
    1    GeV      & 100 000  & 4   \\
    1    TeV      & 100 000 000  & 8  \\
    20   TeV      & 2000 000 000 & 8  \\
   \hline
   \end{tabular}
\caption{The default ProMC mapping between energy units and  C++/JAVA integer representation when using the {\tt int64} varint type of the {\sc Protocol Buffers} library, together with the number of bytes used for the encoding.}
\label{xtab2}
\end{table}

\begin{table}
    \begin{tabular}{llc}
    \hline
    Energy (Gev) &  int representation  & Rounding error (in \%)  \\ \hline

0.0001  &  10   &      10 \\
0.001   & 100    &    1 \\
0.01    & 1,000  &     0.1 \\
0.1     & 10,000 &     0.01 \\
1       & 100,000 &    0.001 \\
10      & 1,000,000  &   0.0001 \\
100     & 10,000,000 &  0.00001 \\
1,000   &  100,000,000 &   0.000001 \\
10,000  &  1,000,000,000 & 0.0000001 \\
21,474  &  2,147,483,647 & 0.00000005 \\
   \hline
    \end{tabular}
\caption{
The mapping between energy values (in GeV) and C++/JAVA integer representation using
the {\tt int64} varint type of the {\sc Protocol Buffers} library and
the multiplicative factor $10^5$ in  the default ProMC varint conversion.
The last column shows the relative rounding errors for the varint conversion.}
\label{xtab2a}
\end{table}

\begin{table}
    \begin{tabular}{llc}
    \hline
    Energy (Gev) &  int representation  & Rounding error (in \%)  \\ \hline
0.001 &   10   &      10 \\
0.01  &   100  &      1 \\
0.1   &   1,000  &      0.1 \\
1     &   10,000  &    0.01 \\
10    &   100,000 &    0.001 \\
100   &   1,000,000 &    0.0001 \\
1,000 &   10,000,000 &  0.00001 \\
10,000 &  100,000,000 &  0.000001 \\
100,000 &  1,000,000,000  & 0.0000001 \\
214,744 & 2,147,483,647 & 0.00000005 \\
\hline
    \end{tabular}
\caption{
Suggested mapping between energy values (in GeV) and  C++/JAVA integer  representation using
the {\tt int64} varint type of the {\sc Protocol Buffers} library
and the multiplicative factor $10^4$ for colliders  with the center-of-mass energies
above $\sqrt{s}=20$  but below $\sqrt{s}=200-400$ TeV.
The last column lists the relative rounding errors for the varint conversion.}
\label{xtab2b}
\end{table}

\begin{table}
   \begin{minipage}{\textwidth}
   \renewcommand{\thefootnote}{\thempfootnote}
   \begin{tabular}{l|p{ 7 cm}}
   \hline
   \multicolumn{2}{ c } {ProMC Tools} \\
   \hline
   {\tt promc\_browser <file>\smallskip} & a JAVA browser to open ProMC files in order to study data layout of data, as well as the stored data.
Currently, the latter feature supports only truth MC records. Files can be streamed using the network. \\
   {\tt promc\_browser\_nlo <file>\smallskip} & a similar browser to open ProMC files with NLO predictions. \\ 
   {\tt promc\_code\smallskip} & Generates analysis code in C++, JAVA, and PYTHON. \\
   {\tt promc\_dump <file>\smallskip} & Prints event numbers, file description, statistics, and meta data to screen. \\
   {\tt promc\_extract <file> <out> N\smallskip} & Extracts a desired number (N) of sequential events and saves them to another file. \\
   {\tt promc\_info <file>} & Displays the information of the ProMC file. \\
   {\tt promc\_log <file>\smallskip} & Extracts a log file (if attached). \\
   {\tt promc\_proto <file>\smallskip} & Extracts the file layouts in the form of Protocol Buffers data templates. \\
   {\tt promc\_split <file> N} & Splits a ProMC file into desired number (N) of smaller files\footnote{To be compiled separately since depends on the actual data structure.} \\
   \hline
   \end{tabular}
   \caption{A list of tools and commands available in the {\sc ProMC} package. The tools are can be called from any file path upon set up of {\sc ProMC}.}
   \label{xtab3}
   \end{minipage}
\end{table}

\begin{table}
   \begin{minipage}{\textwidth}
   \renewcommand{\thempfootnote}{\arabic{footnote}}
   \addtocounter{footnote}{+1}%
   \begin{tabular}{l|p{ 10 cm}}
   \hline
   \multicolumn{2}{ c } {ProMC conversion tools} \\
   \hline
   {\tt hepmc2promc} & Converts a HEPMC 2.03.11 file \cite{hepmc} to the {\sc ProMC} file format. \\
   {\tt promc2hepmc} & Converts a {\sc ProMC} file to a HEPMC 2.03.11 file \cite{hepmc}. \\
   {\tt promc2root} & Converts stores a {\sc ProMC} file in a ROOT tree \cite{root}. \\
   {\tt stdhep2promc} & converts a STDHEP file \cite{stdhep} to the {\sc ProMC} file. \\
   \hline
   \end{tabular}
   \caption{A list of converters from/to different file formats supported by the {\sc ProMC} library. The tools should are located in the directory "examples" of the {\sc ProMC} packages and should be compiled.}
   \label{xtab4}
   \end{minipage}
\end{table}

\end{document}